# High Speed Mid-Infrared Interband Cascade Photodetector Based on InAs/GaSb Type-II Superlattice

Yaojiang Chen, Xuliang Chai, Zhiyang Xie, Zhuo Deng, Ningtao Zhang, Yi Zhou, Zhicheng Xu, Jianxin Chen, Baile Chen

*Abstract*—**High speed mid-wave infrared (MWIR) photodetectors have applications in the areas such as free space optical communication and frequency comb spectroscopy. However, most of the research on the MWIR photodetectors is focused on how to increase the quantum efficiency and reduce the dark current, in order to improve the detectivity (D*), and the 3dB bandwidth performance of the corresponding MWIR photodetectors is still not fully studied. In this work, we report and characterize a MWIR interband cascade photodetector based on InAs/GaSb type-II superlattice with a 50% cutoff wavelength at ~5.3 µm at 300 K. The 3 dB cutoff frequency is 2.4 GHz at 300 K, for a 40 µm circular diameter device under -5 V applied bias. Limitations on the detector high speed performance are also discussed.**

## I. Introduction

HIGH performance mid-wave infrared (MWIR) lasers and photodetectors are essential components of the recent developed high speed photonics application such as free space optical communication system and frequency comb spectroscopy. The utilization of free space optical communication system is expected to be able to overcome the issue of bandwidth, spectrum and security of the radio frequency (RF) communication system [1]–[3]. In atmospheric environment, the propagation of free space light will benefit from the minimum scattering and attenuation at mid/long wave infrared band [1], [3]. In recent years, quantum cascade laser (QCL) and interband cascade laser (ICL) operating at 3-5 µm wavelength have brought prospective light sources to the free space optical communication system at this band [4]–[6]. On the other hand, newly developed frequency comb spectroscopy techniques are attractive for its much higher precision than traditional spectroscopy system [7]–[9]. For MWIR band where characteristic spectrum of most gas molecules locate at, the mode-lock QCL and ICL have also been developed as the light sources for frequency comb [10], [11]. Meanwhile, both of these applications require high speed photodetectors in the MWIR band, which are currently not fully developed. Photodetectors based on mercury cadmium telluride need cooling [12], [13] and are not compatible with the global initiative to eventually phase out the use of mercury [14]. Although the quantum well infrared photodetector can achieve high response speed, it has shortcomings in absorption efficiency and complex grating systems [15], [16]. The tunability on cutoff wavelength and excellent carrier transport properties have made InAs/GaSb type-II superlattice (T2SL) a preferable material for MWIR detection [17]. The interband cascade infrared photodetector (ICIP) based on InAs/GaSb T2SL operating at 3-5 µm wavelength shows superiorities on operation temperature, carrier collection efficiency and signal to noise ratio (SNR) [18]. Nevertheless, the high-speed performance of ICIP is not fully studied. Taking the advantage of multiple cascade stage architecture, the transit time of photogenerated carriers can be shortened without sacrificing the sensitivity, thus the theoretical response speed of ICIP will be higher than the other InAs/GaSb photodetectors with similar thickness of absorber. Lotfi et al. reported an extracted bandwidth of 1.3 GHz for a three-stage ICIP, and concluded that the bandwidth can be improved by reducing parasitic circuit parameters [19].

In this work, we demonstrate a two-stage ICIP based on InAs/GaSb type-II superlattice and the speed response performances of the device are extensively characterized. The 3dB bandwidth ($f_{3dB}$) of a 40 µm circular diameter photodiode is 2.4 GHz at 300 K under -5 V bias voltage. An equivalent circuit model is used to analyze the RC limit of the device, and results show that the response speed is limited by carrier transit time.

This work was supported by the Shanghai Sailing Program (17YF1429300), the National Key Research and Development Program of China (No. 2018YFB2201000)) and ShanghaiTech University startup funding (F-0203-16-002).

Yaojiang Chen, Zhiyang Xie, Zhuo Deng, Ningtao Zhang and Baile Chen are with the School of Information Science and Technology, ShanghaiTech University, Shanghai 201210, China. Yaojiang Chen is also with the Shanghai Institute of Microsystem and Information Technology, Chinese Academy of Sciences, Shanghai 200050, China, and also with University of Chinese Academy of Sciences, Beijing 100049, China (corresponding author to provide e-mail: chenbl@shanghaitech.edu.cn).

Xuliang Chai, Yi Zhou, Zhicheng Xu and Jianxin Chen are with the Key Laboratory of Infrared Imaging Materials and Detector, Shanghai Institute of Technical Physics, Chinese Academy of Sciences, Shanghai 200083, China. Xuliang Chai is also with University of Chinese Academy of Sciences, Beijing 100049, China (corresponding authors to provide e-mail: zhouyi@mail.sitp.ac.cn).

Yaojiang Chen and Xuliang Chai contribute equally to the work.

## II. Device Structure

The epitaxial structure of the two-stage ICIP is shown in Fig. 1. The absorber consists of 1250 nm and 1110 nm InAs/GaSb superlattice for each of the two stages respectively. The photo-generated electrons in the absorber will transport towards n-side through the carefully designed InAs/AlSb superlattice by phonon assisted tunneling, while the photo-generated holes transport towards p-side through the AlSb/GaSb superlattice. Meanwhile, InAs/GaSb and AlSb/GaSb superlattice are also used as the hole barrier and electron barrier respectively to block the carrier to diffuse to the opposite direction. The whole structure is properly designed so that the photo-generated electrons can gradually relax through hole barrier and recombine with the holes from the adjacent electron barrier region.

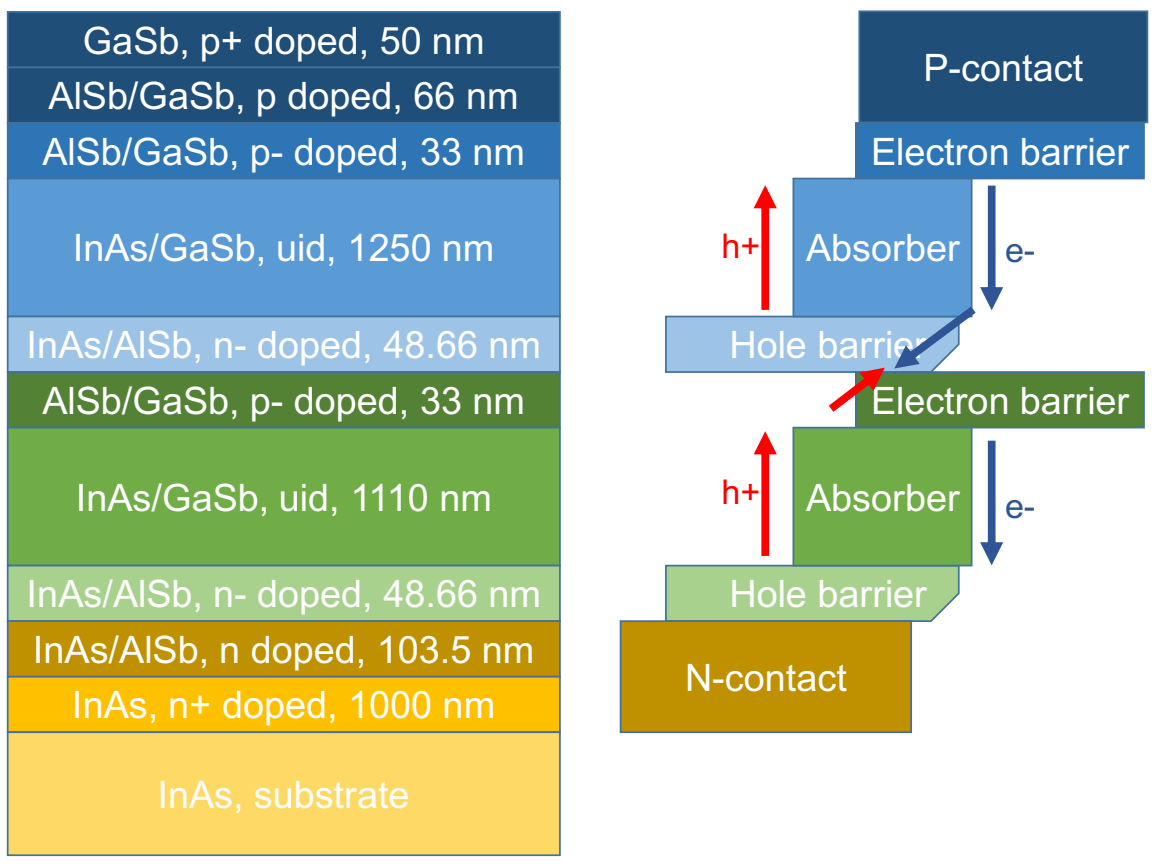

Fig. 1. Epitaxial structure of the two-stage ICIP. The schematic on the right shows the band diagram and the diffusion/drift of electrons (e-) and holes (h+).

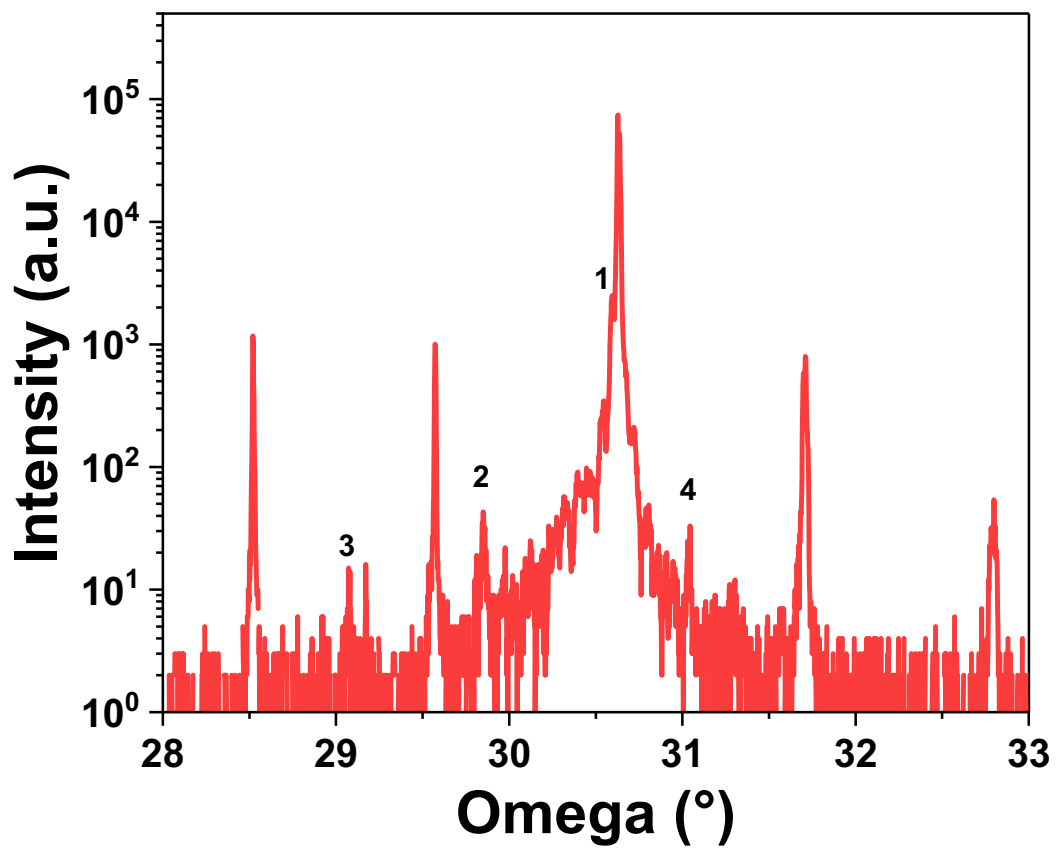

Fig. 2. High resolution X-ray diffraction pattern of the structure.

The crystal quality was investigated by the high-resolution X-ray diffraction pattern as shown in Fig. 2. A lattice mismatch of $1.35\times10^{-4}$ is found between superlattice and the substrate. The full-width-at-half-maximum (FWHM) of the SL(-1) peak is 23.47 arcsec. The calculated period thickness of absorption region is 4.82 nm which is in very good agreement with the designed value of 4.85 nm. Peak 1 in Fig. 2. is the SL(0) peak of InAs/AlSb superlattice n-contact layer and peak 4 is the corresponding SL(+1) peak. The fitted period thickness is the same as the designed value of 9.3 nm. Peak 2 and 3 are satellite peak of AlSb/GaSb electron barrier, and the fitted period thickness is 6.6 nm, which also agrees with the designed value.

## III. DC Performance and Optical Response

For DC performance and optical response measurement, the wafer was processed into mesa type devices with different diameter using standard contact lithography and wet etching. The devices were passivated by 300 nm SiN layer. Ti/Pt/Au were deposited for ohmic contact. Device with 300 µm diameter were mounted into Dewars with wire-bonding for DC electrical property and optical response measurement.

The dark current and dark current density of the photodiode was measured at various temperature, as shown in Fig. 3 (a). The dark current density is $2.93\times10^{-4}$ A/cm$^2$ at 148 K under -0.1 V bias and increases to 3.97 A/cm$^2$ at 304 K. As shown in Fig. 3 (b), linear fit of the Arrhenius plot at -0.1 V shows an activation energy ($E_a$) of 239.7 meV, which is almost equal to the effective bandgap of the InAs/GaSb absorber (239.7 meV~243 meV) and indicating a diffusion-dominated dark current.

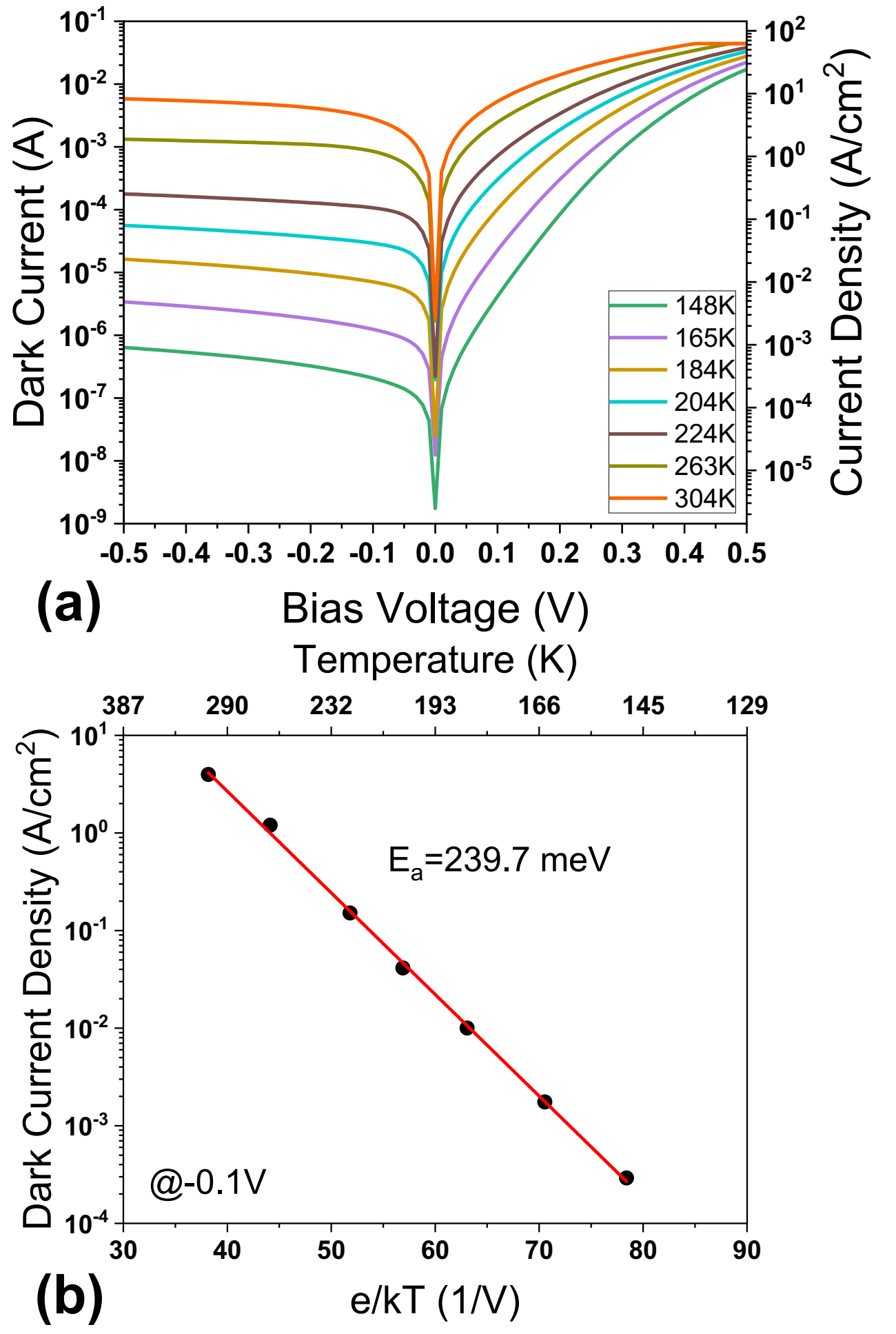

Fig. 3. (a) Dark current and dark current density versus the bias voltage measured from the two-stage ICIP at various temperatures from 148 K to 304 K. (b) Arrhenius plot of the dark current density at -0.1 V.

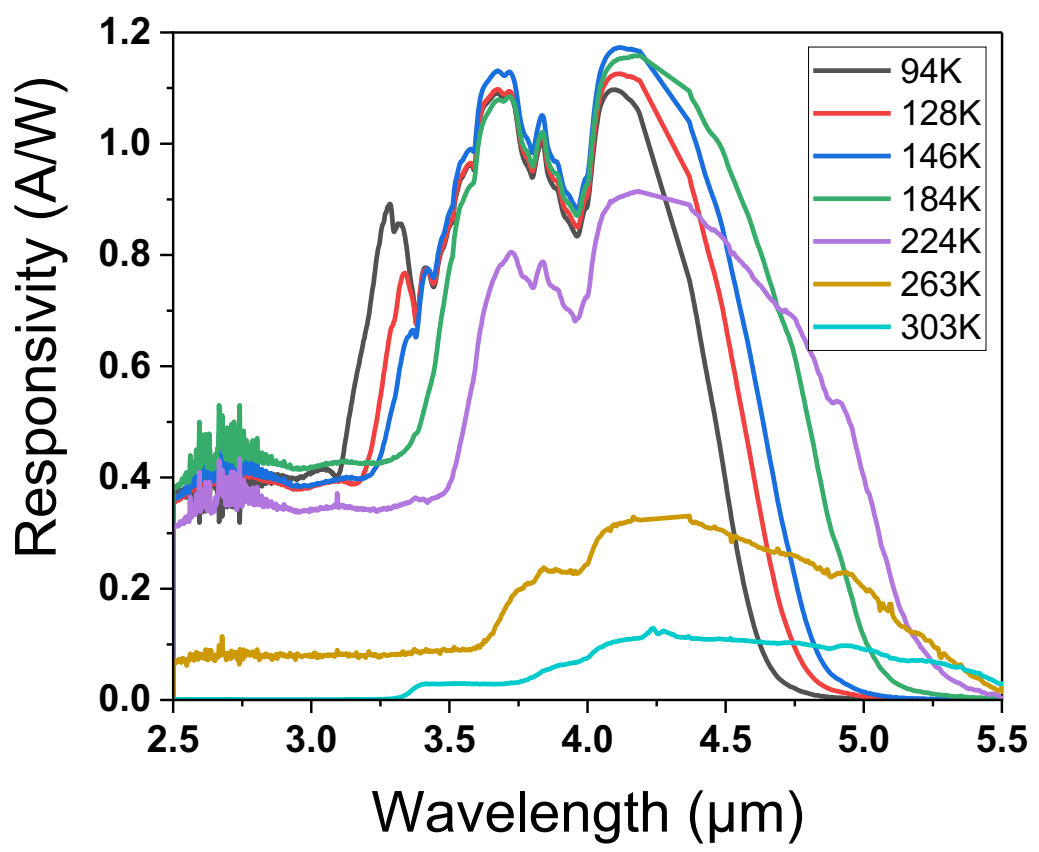

Fig. 4. Responsivity of the two-stage ICIP measured from 94 K to 303 K under 0V bias voltage.

The responsivity spectra of the two-stage ICIP at various temperatures at 0 V bias are shown in Fig. 4. A peak responsivity of 1.17 A/W can be observed at 146 K and 4.1 μm. The 50% cutoff wavelength locates at ~4.4 μm at 94 K. When temperature increases to 303 K, the cutoff wavelength monotonically red-shifts to ~5.3 μm. The peak responsivity exhibits a slight increase as the temperature increases to 184 K. However, it drops significantly at higher temperature (>184 K) due to the increased carrier recombination rate and decreased carrier life time and diffusion length [17], [20].

## IV. Bandwidth Characterization

For 3dB bandwidth characterizations, a 40 μm diameter normal incident photodiode was fabricated by wet etching. The electrodes were deposited by e-beam evaporation with Ti/Pt/Au and Ge/Au/Ni/Au for p-contact and n-contact respectively. Afterwards, a coplanar waveguide pad with 50 Ohm characteristic impedance was electroplated on 2 um thickness SU-8 for high speed RF measurement. Before the frequency response measurement, the dark current of the 40 μm diameter device was also measured at 300K and 200K for verification as shown in Fig. 5. The dark current density of the device at high temperature is very close to device with large diameter as shown in Fig. 3 (a).

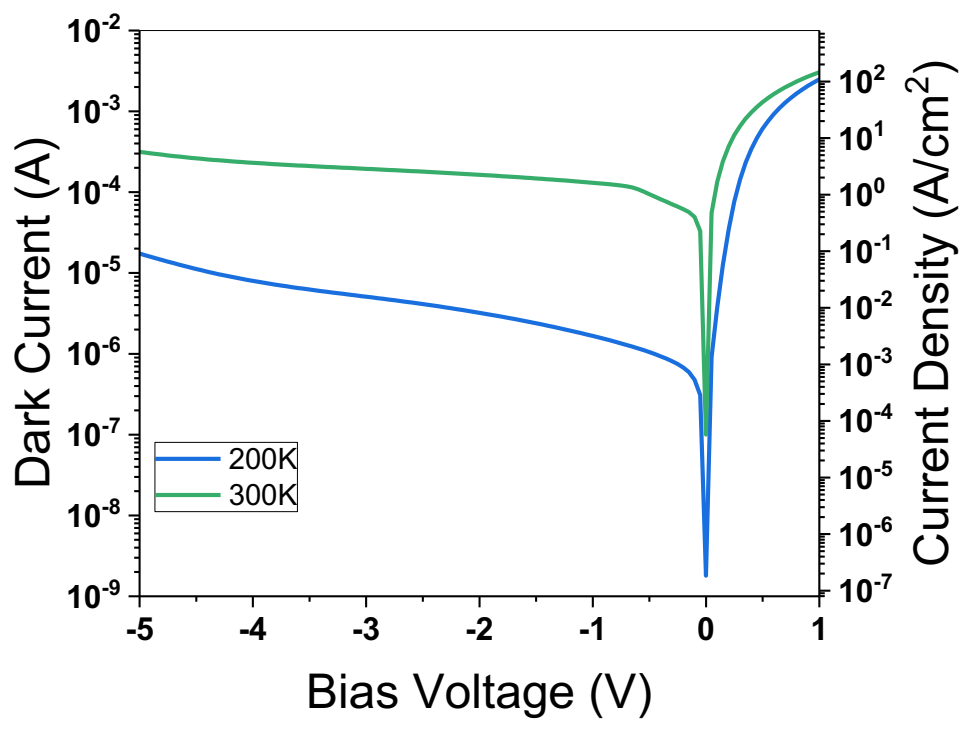

Fig. 5. Dark current and dark current density versus the bias voltage measured from the 40 μm diameter device at various temperatures from 200 K to 300 K.

The frequency response of the photodiode was investigated by a calibrated lightwave component analyzer (LCA). An intensity-modulated light of 1550 nm wavelength was generated by the LCA. The light was focused by a lens fiber and coupled into the photodiode. The photoresponse of the photodiode was gathered by a GSG probe. Then the RF component of the photoresponse was separated by a bias tee and returned to the LCA to compare with the original signal. The bias of the photodiode was provided by a source meter through the DC port of the bias tee.

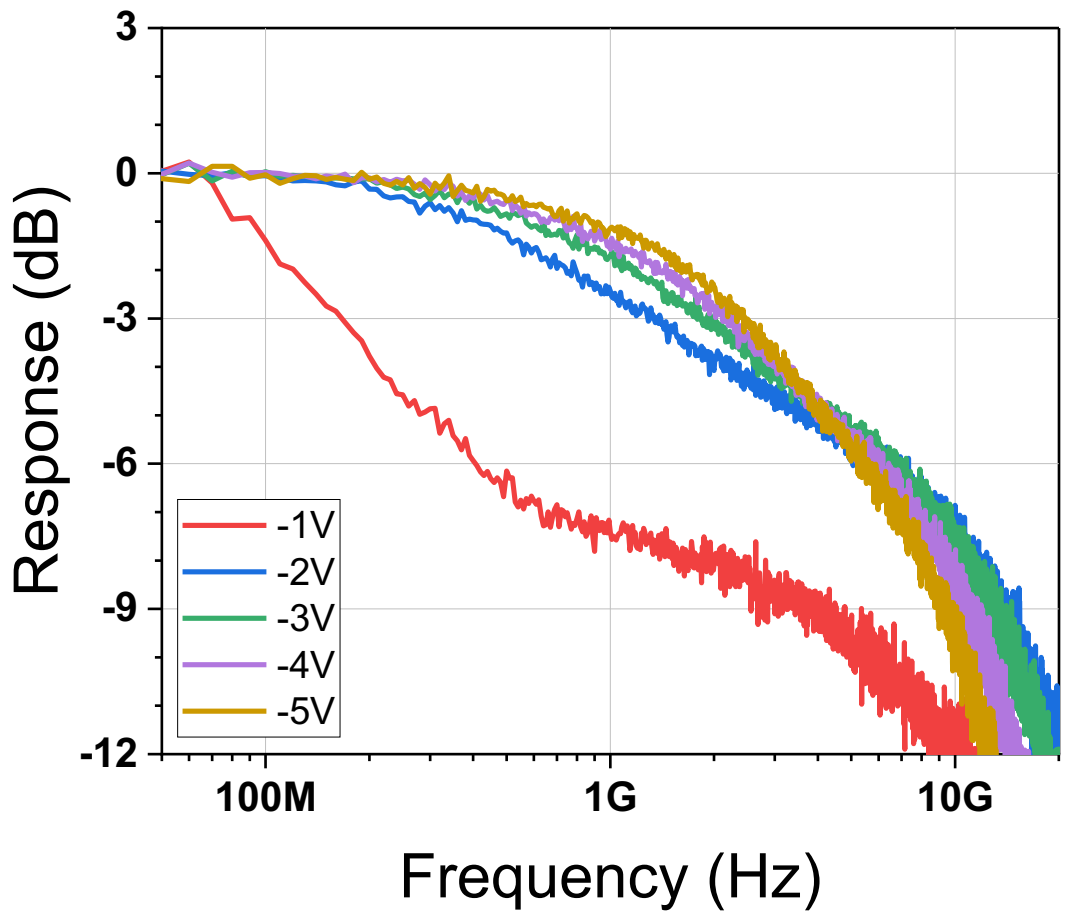

Fig. 6. Frequency response of the two-stage ICIP with 40 μm diameter at 300 K and various bias voltages.

The frequency responses under various bias voltages (-1 V to -5 V) of the two-stage ICIP with 40 μm diameter were measured at 300 K, as shown in Fig. 6. The 3dB bandwidth increases from 170 MHz to 2.4 GHz when the bias voltage increases from -1 V to -5 V. This phenomenon can be explained by the faster carrier transport velocity induced by stronger electrical field at larger reverse bias. The 3dB bandwidth should be larger at higher bias voltage, but the leakage current also rises and may damage the photodiode. Fig. 7 (a) plots the temperature dependences of frequency response under -5 V applied bias. When temperature varies, the 3dB bandwidth shows a subtle change and has a peak value of 2.6 GHz at 200 K. As temperature decreases from 300K to 200K, it is expected the diffusion length of the photo-generated carrier increases, and the transit time would decrease, thus the 3dB bandwidth increases, which is also consistent to temperature varied responsivity measurement. When the temperature further decreases from 200K to 77K, it is expected the phonon assisted tunneling in the AlSb/GaSb electron barrier and InAs/AlSb hole barrier layers would be less efficient, and slow down the transport of photo-generated carriers, therefore the bandwidth decreases. That is why the 3dB bandwidth shows a peak value around 200K. Fig. 7 (b) summaries the 3dB bandwidth transition with respect to the temperature under different bias voltages. For each bias voltage, the 3dB bandwidth firstly increases with temperature (< 200 K) and then exhibits slight decrease at higher temperature (> 200 K). And for each temperature, the 3dB bandwidth increases with the increase of reverse bias voltage.

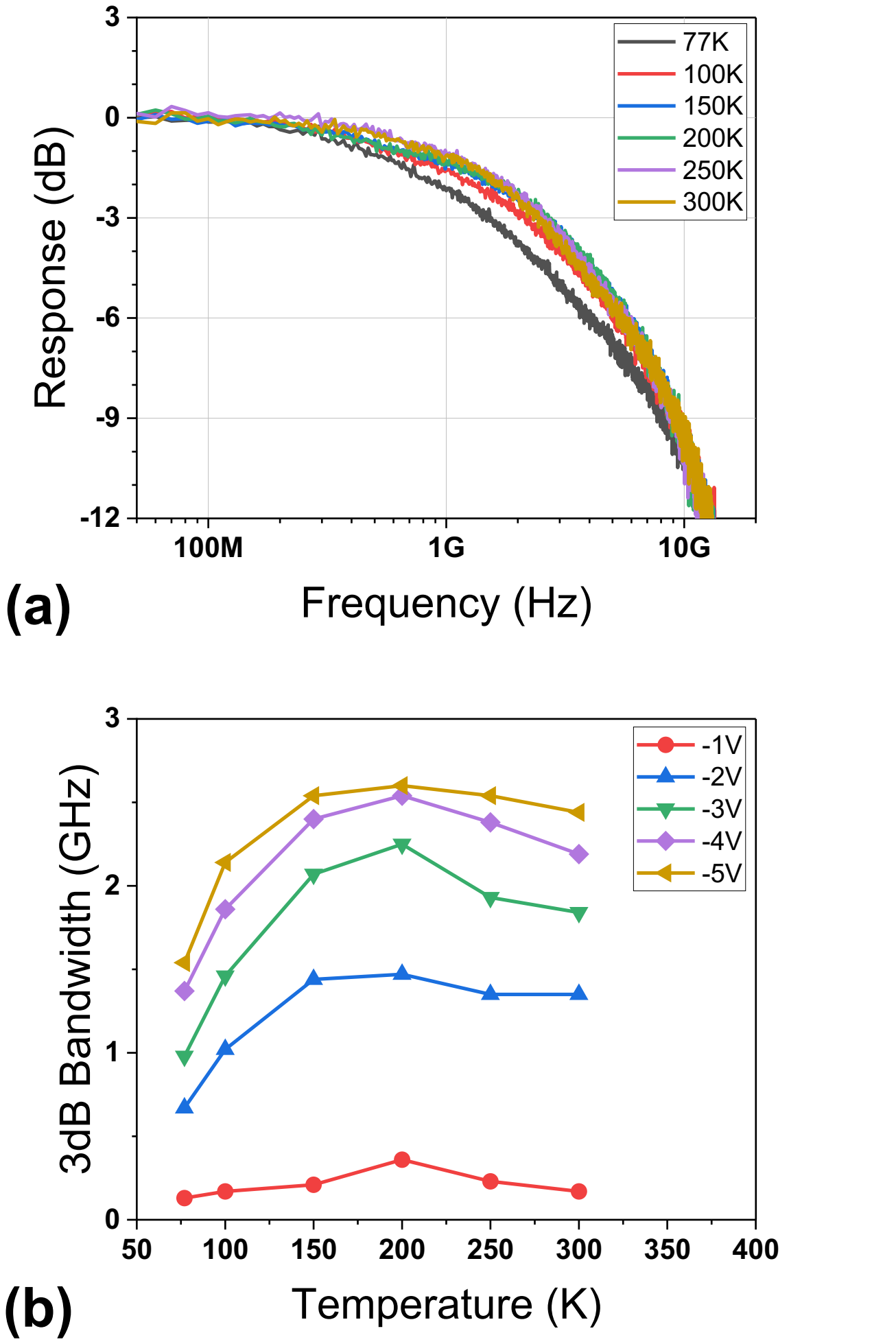

Fig. 7. (a) Dependence of frequency response on temperature of the 40 µm two-stage ICIP at -5 V. (b) 3dB bandwidth transition of the 40 µm two-stage ICIP with respect to temperature at different bias voltage.

To further investigate the mechanism of the high-speed characteristics of the two-stage ICIP, the electrical S11 parameter under various bias voltages in dark environment was measured, as shown in Fig. 8 (a-c), and fitted with an equivalent circuit model in Fig. 8 (d). At reverse bias, the two-stage ICIP can be regarded as two p-i-n junctions connecting in series, while each junction consists of a junction resistance ($R_{j1}$ or $R_{j2}$) in parallel with a junction capacitance ($C_{j1}$ or $C_{j2}$). $R_s$ includes the contact resistance and the series resistance of the electrical circuit. $C_p$ is the parasitic capacitance of the coplanar waveguide pad.

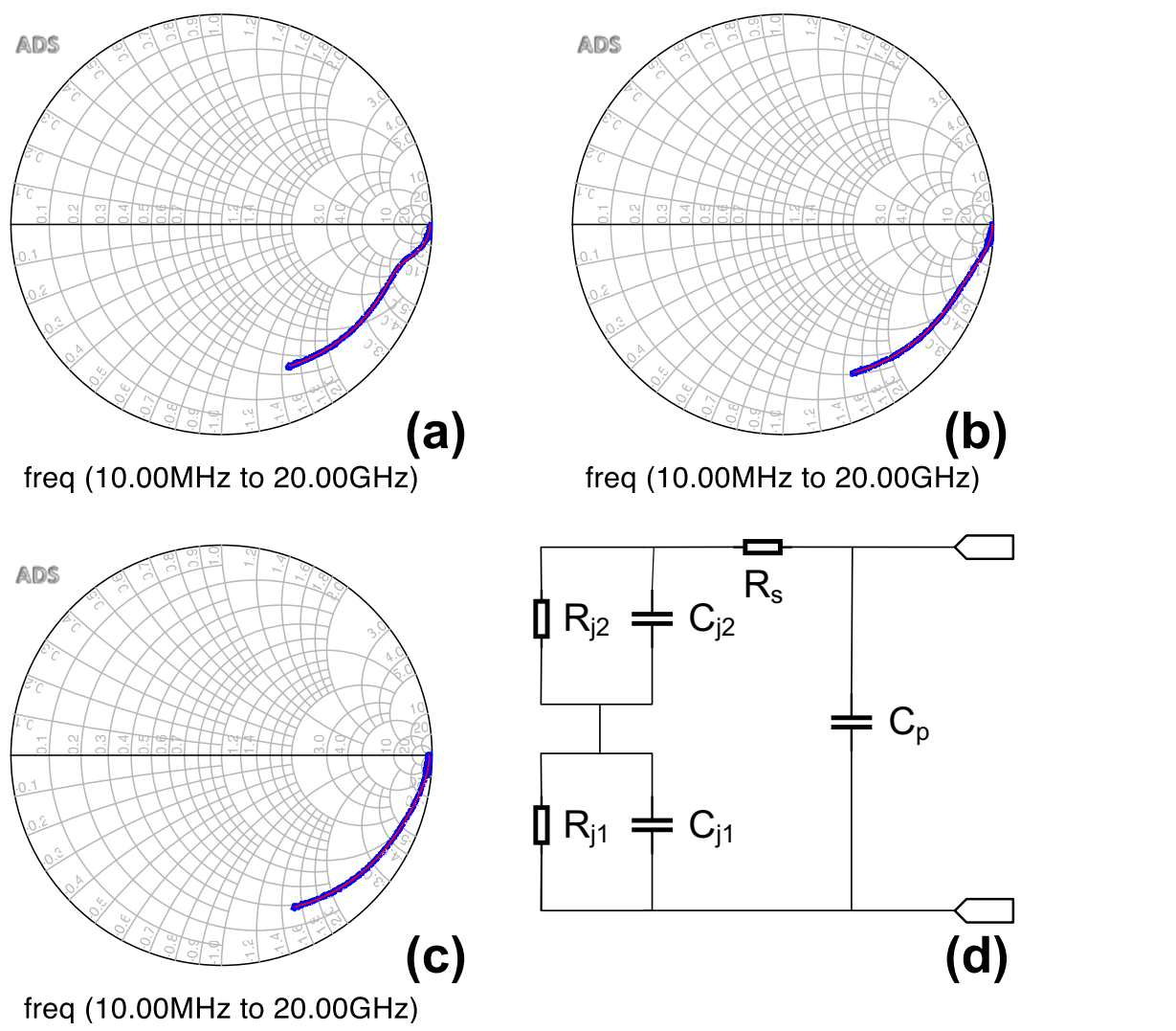

Fig. 8. S11 parameter under (a) -1 V, (b) -3 V, and (c) -5 V bias voltage at 300 K. (d) Equivalent circuit model of the two-stage ICIP.

The fitting results of the model parameters under various bias voltages at 300 K are shown in Fig. 9. In Fig. 9 (a), the junction capacitance of stage 1 ($C_{j1}$) decreases with the increase of reverse bias voltage, which is in consistent with the capacitance transition of a theoretical p-i-n junction under reverse bias. The junction resistance of stage 1 ($R_{j1}$) increases first and then decreases with the bias voltage, as shown in Fig 9 (b). The increase of $R_{j1}$ from -1 V to -3 V is due to the reduction of carriers in depletion region, and the further decrease at large bias (-3 V to -5 V) can be attributed to the enhanced tunneling current. The transition of $C_{j1}$ and $R_{j1}$ indicates that stage 1 is gradually depleted with increasing bias voltage. For stage 2, the values of $C_{j2}$ and $R_{j2}$ show only weak dependences on reverse bias, indicating that the electric field drops in stage 2 is small. We speculate stage 2 is far from fully depletion. Based on the equivalence circuit modeling results, it is believed that there is a significantly asymmetrical bias drop in these two stages, in order to maintain the current matching condition [21]. Hence, only one stage shows the characteristic of a p-i-n junction. The absence of electric field in the un-depleted absorber of the other stage may post limitation on the carrier transit process, which could affect the response speed of the photodiode. The values of $R_s$ and $C_p$ can be regarded as constant because they are theoretically independent of the bias voltage.

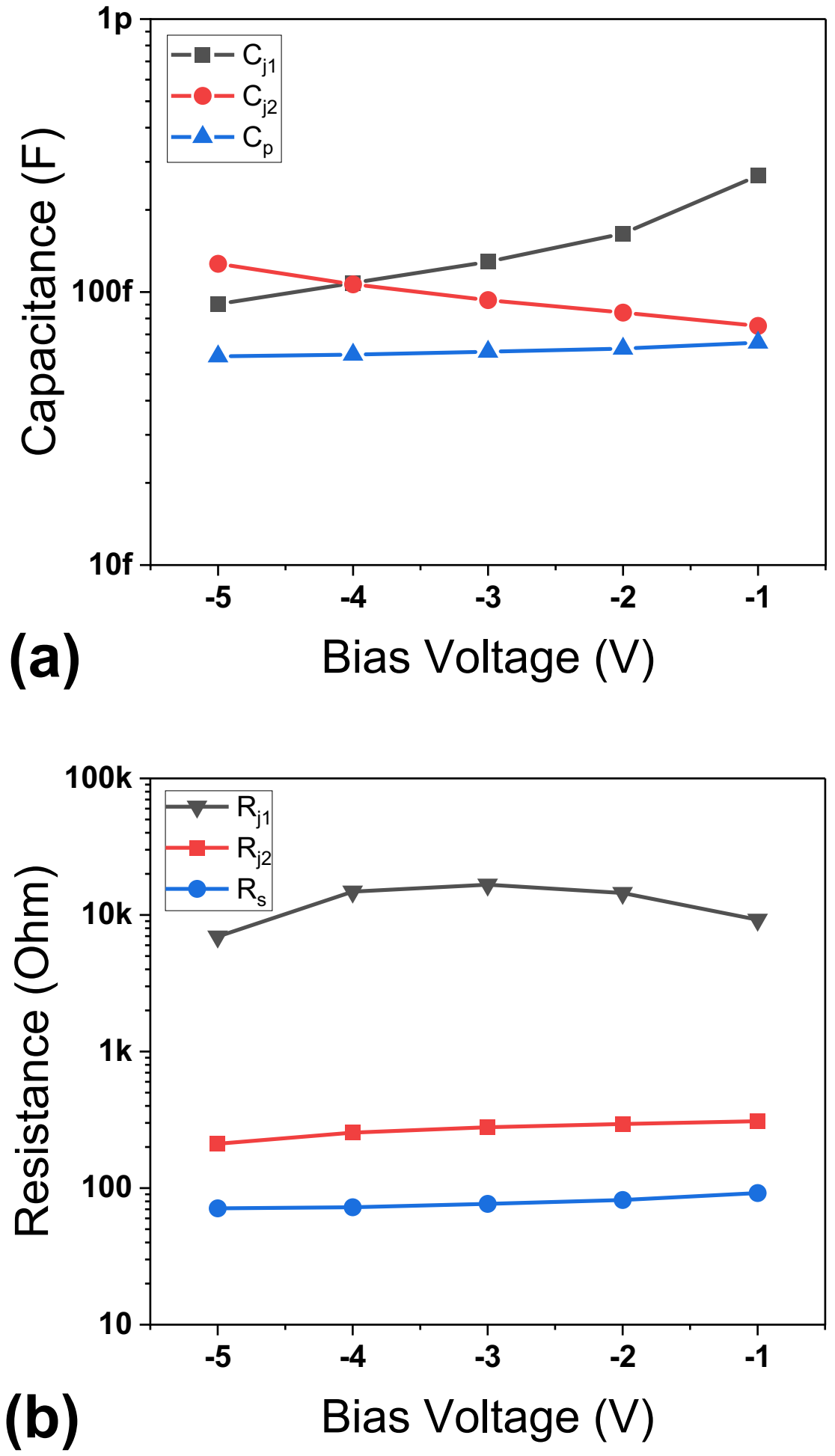

Fig. 9. Dependence of extracted circuit parameters on bias voltage of the 40 μm two-stage ICIP at 300 K.

According to the extracted model parameters, the RC limit bandwidth $f_{RC}$ can be simulated by circuit simulation software (Advanced Design System). Fig. 10 (a) shows the frequency response of the equivalent circuit model using extracted parameters. RC limit bandwidth of 16.9 GHz and 25.1 GHz can be found at 300 K and 200 K under -5 V, respectively. Fig. 10 (b) shows the bias voltage dependences of the calculated RC limit bandwidth and the 3dB bandwidth of the two-stage ICIP at different temperatures. It is clear that the value of $f_{RC}$ is much larger than $f_{3dB}$, indicating that the overall device bandwidth is limited by the carrier transit process. Considering that one stage is not fully depleted, we believe the transit time is limited by the slow diffusion process in superlattice absorber of the un-depleted stage. It is also noted that the 3dB bandwidth of the devices is characterized with 1550nm laser, which is almost fully absorbed in the first stage. This asymmetric absorption profile in the two stages will cause current matching issue in the ICIP[21], which requires additional bias to maintain the current matching condition. One of our future work will be using the high speed MWIR light source to characterize these photodetectors, which could enable more symmetric absorption profile in both stages, and acquire less operational bias or even zero bias for high speed application. To further improve the speed of the photodetectors, multiple stages with thinner absorber would also be desirable.

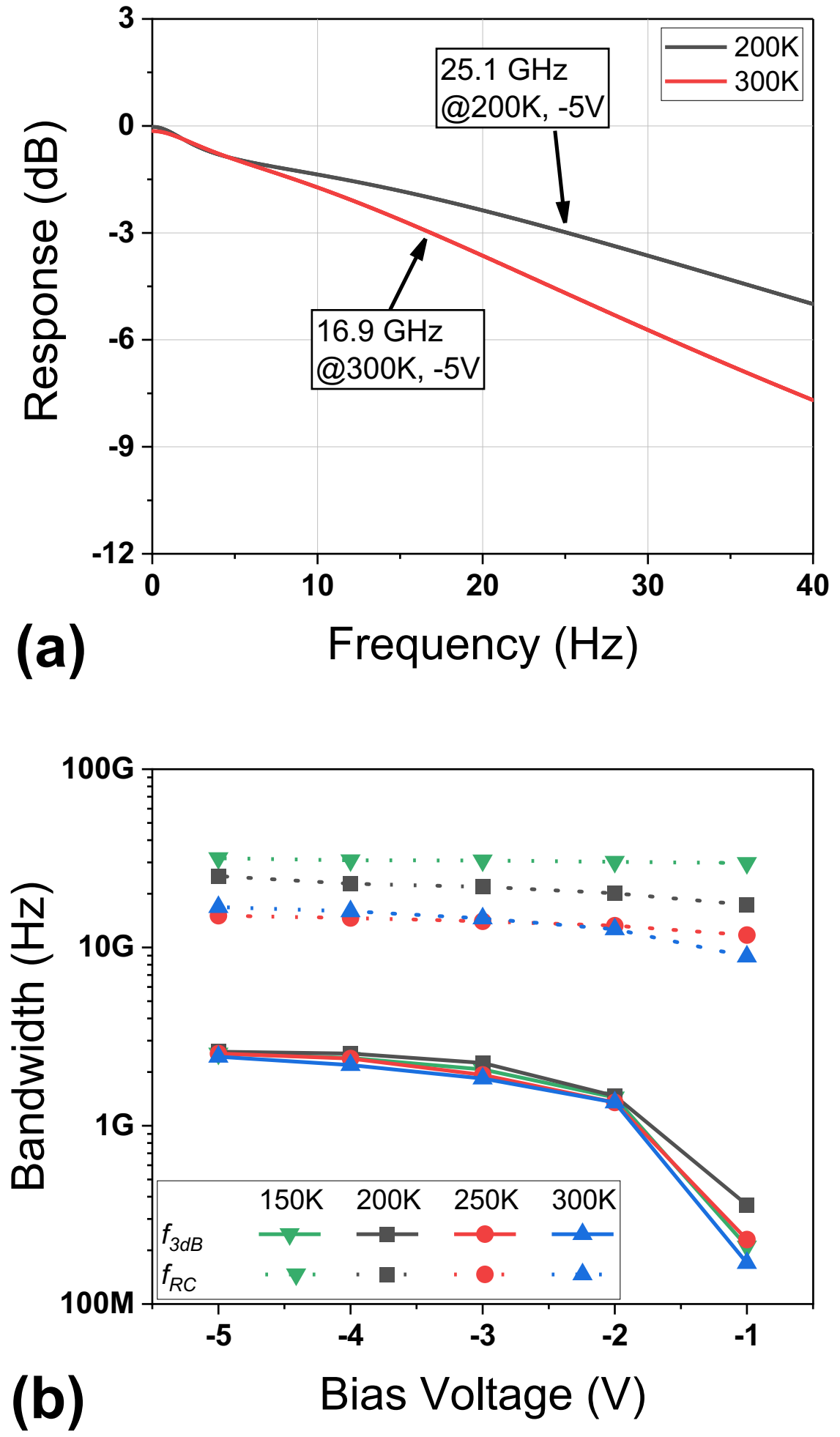

Fig. 10. (a) Frequency response of the extracted equivalent circuit model of the two-stage ICIP at 200 K and 300 K under -5 V bias voltage. (b) Bias voltage dependences of RC limit bandwidth (dash lines) and 3dB bandwidth (solid lines) of the 40 μm two-stage ICIP at different temperatures.

## V. Conclusion

In summary, we have demonstrated a two-stage ICIP based on InAs/GaSb type-II superlattice for high speed MWIR detection, and the frequency response characteristics have been extensively studied. The 3dB bandwidth of the demonstrated photodetector can achieve 2.4 GHz at 300 K under -5 V bias voltage. According to the fitting results of S11 parameters, the overall device bandwidth is mainly limited by the slow carrier transit in un-depleted superlattice absorber in one of the stages. Utilizing high speed MWIR light source will create more symmetric absorption profile in both stages, and our future work will be focused on that.

## References

[1] H. Henniger and O. Wilfert, "An introduction to free-space optical communications," *Radioengineering*, vol. 19, no. 2, pp. 203–212, 2010.

[2] V. W. S. Chan, "Free-Space Optical Communications," vol. 24, no. 12, pp. 4750–4762, 2006.

[3] M. A. Khalighi and M. Uysal, “Survey on free space optical communication: A communication theory perspective,” *IEEE Commun. Surv. Tutorials*, vol. 16, no. 4, pp. 2231–2258, 2014.

[4] A. Delga and L. Leviandier, “Free-space optical communications with quantum cascade lasers,” in *Proc.SPIE*, 2019, vol. 10926.

[5] C. Liu *et al.*, “Free-space communication based on quantum cascade laser,” *J. Semicond.*, vol. 36, no. 9, 2015.

[6] A. Soibel *et al.*, “Midinfrared Interband Cascade Laser for Free Space Optical Communication,” *Free. Laser Commun. Technol. XXII*, vol. 7587, no. 2, p. 75870S, 2010.

[7] I. Coddington, W. C. Swann, and N. R. Newbury, “Coherent multiheterodyne spectroscopy using stabilized optical frequency combs,” *Phys. Rev. Lett.*, vol. 100, no. 1, pp. 11–14, 2008.

[8] I. Coddington, N. Newbury, and W. Swann, “Dual-comb spectroscopy,” *Optica*, vol. 3, no. 4, pp. 414–426, 2016.

[9] F. C. Cruz *et al.*, “Mid-infrared optical frequency combs based on difference frequency generation for molecular spectroscopy,” *Opt. Express*, vol. 23, no. 20, p. 26814, 2015.

[10] A. Hugi, G. Villares, S. Blaser, H. C. Liu, and J. Faist, “Mid-infrared frequency comb based on a quantum cascade laser,” *Nature*, vol. 492, no. 7428, pp. 229–233, 2012.

[11] M. Bagheri *et al.*, “Passively mode-locked interband cascade optical frequency combs,” *Sci. Rep.*, vol. 8, no. 1, pp. 1–7, 2018.

[12] C. T. Elliott, N. T. Gordon, and A. M. White, “Towards background-limited, room-temperature, infrared photon detectors in the 3-13 μm wavelength range,” *Appl. Phys. Lett.*, vol. 74, no. 19, pp. 2881–2883, 1999.

[13] A. Rogalski, “HgCdTe infrared detector material: History, status and outlook,” *Reports Prog. Phys.*, vol. 68, no. 10, pp. 2267–2336, 2005.

[14] Minamata Convention, “Minamata Convention on Mercury Text and Annexes,” *Minamata Convention*, 2017. [Online]. Available: http://www.mercuryconvention.org/. [Accessed: 27-Apr-2018].

[15] D. Palaferri *et al.*, “Room-temperature nine-μm-wavelength photodetectors and GHz-frequency heterodyne receivers,” *Nature*, vol. 556, p. 85, Mar. 2018.

[16] E. Rodriguez *et al.*, “Room temperature, wide-band Quantum Well Infrared Photodetector for microwave optical links at 4.9 μm wavelength,” *ACS Photonics*, p. acsphotonics.8b00704, 2018.

[17] Z. Deng, D. Guo, J. Huang, H. Liu, J. Wu, and B. Chen, “Mid-wave infrared InAs/GaSb type-II superlattice photodetector with n-B-p design grown on GaAs substrate,” *IEEE J. Quantum Electron.*, vol. 55, no. 4, pp. 1–5, 2019.

[18] W. Huang *et al.*, “Electrical gain in interband cascade infrared photodetectors,” vol. 113104, 2018.

[19] H. Lotfi *et al.*, “High-frequency operation of a mid-infrared interband cascade system at room temperature,” *Appl. Phys. Lett.*, vol. 108, no. 20, pp. 1–5, 2016.

[20] R. T. Hinkey and R. Q. Yang, “Theory of multiple-stage interband photovoltaic devices and ultimate performance limit comparison of multiple-stage and single-stage interband infrared detectors,” *J. Appl. Phys.*, vol. 114, no. 10, 2013.

[21] W. Huang *et al.*, “Current-matching versus non-current-matching in long wavelength interband cascade infrared photodetectors,” *J. Appl. Phys.*, vol. 122, no. 8, 2017.